\documentclass[prb,floats,twocolumn]{revtex4}
\usepackage{graphicx}

\usepackage{amsfonts}      % Paquete AMS
\usepackage{amssymb}      % Paquete AMS
\usepackage{amsbsy}      % Paquete AMS
\usepackage{amsmath}

\begin{document}

\title{One dimensionalization in the spin-$1$ Heisenberg model on the anisotropic triangular lattice}
\author{M. G. Gonzalez, E. A. Ghioldi, C. J. Gazza, L. O. Manuel, and A. E. Trumper}
\affiliation {Instituto de F\'isica Rosario, CONICET - Universidad Nacional de
Rosario, Boulevard 27 de Febrero 210 Bis, 2000 Rosario, Argentina }

\begin{abstract}
We investigate the effect of dimensional crossover in the ground state of the antiferromagnetic spin-$1$ Heisenberg model on the anisotropic triangular lattice that interpolates between the regime of weakly coupled Haldane chains ($J^{\prime}\! \!\ll\!\! J$) and the isotropic triangular lattice ($J^{\prime}\!\!=\!\!J$). We use the density-matrix renormalization group (DMRG) and Schwinger boson theory performed at the Gaussian correction level above the saddle-point solution. Our DMRG results show an abrupt transition between decoupled spin chains and the spirally ordered 
regime at $(J^{\prime}/J)_c\sim 0.42$, signaled by the sudden closing of the spin gap. Coming from the magnetically ordered side, the computation of the spin stiffness within Schwinger boson theory predicts the instability of the spiral magnetic order toward a magnetically disordered phase with one-dimensional features at $(J^{\prime}/J)_c \sim 0.43$. The agreement of these complementary methods, along with the strong difference found between the intra- and the interchain DMRG short spin-spin correlations; for sufficiently large values of the interchain coupling, suggests that the interplay between the quantum fluctuations and the dimensional crossover effects gives rise to the one-dimensionalization phenomenon in this frustrated spin-$1$ Hamiltonian.
\end{abstract}
\maketitle

\section{Introduction}

The role played by quantum fluctuations in low dimensional antiferromagnets 
is quite well understood when frustration is not present. \cite{Auerbach94} For one dimensional (1D) systems the Haldane conjecture,\cite{Haldane83} 
regarding the gapless and gapped magnetic excitations for $s=1/2$ and $s=1$, respectively, has been largely confirmed theoretical and experimentally.\cite{Nagler91}
Here, the magnetic excitation spectra of the critical ($s=1/2$) and the Haldane ($s=1$) phases are successfully interpreted in terms of 
spin-$1/2$ spinons and spin-$1$ triplet excitations, respectively.  In two dimensions, like in the square lattice, the rupture of the $SU(2)$ symmetry of the N\'eel 
ground state has been widely confirmed\cite{Neves86} in both cases, $s=1/2$ and $s=1$, where the magnetic excitations are well described by spin-$1$ magnonic excitations. In the 
spin-$1/2$ case, however, it has been proposed that some high-energy anomalies observed in the spectrum of the cuprates superconductors could be explained 
by the mean of fermionic spinon excitations\cite{cuprates}. \\

\begin{figure}[h]
\begin{center}
\includegraphics*[width=0.35\textwidth]{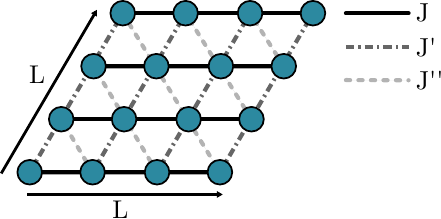}
\caption{ Geometry of the AF exchange interactions. The interpolation among the 
different systems is explained in the text}
\label{red}
\end{center}
\end{figure}

The effect of the dimensional crossover in these systems is also very interesting. In interpolating from decoupled spin chains to the square spin 
lattice (see Fig.\ref{red} with $J=1$, $J^{\prime\prime}=0$, and $J^{\prime}$ varying from $0$ to $1$ ) the behavior of the antiferromagnetic (AF) Heisenberg model depends strongly on the spin value. The critical feature of the spin-$1/2$ chain ground state makes the system susceptible to breaking the $SU(2)$ symmetry with an infinitesimal interchain coupling $J^{\prime}/J$ and develops long range N\'eel order; however for the spin-$1$ case\cite{Matsumoto01}, it takes a very small value $(J^{\prime}/J)_c\sim 0.0436$, notably one order of magnitude smaller than the spin chain gap, to quench the Haldane phase and, simultaneously, to develop long range N\'eel order\cite{Wierschem14}.  \\
   
The interpolation between the decoupled chains and the frustrated 
triangular lattice (see Fig \ref{red} with $J=1$ and $J^{\prime \prime}=J^{\prime}$, with $J^{\prime}$ varying from $0$ to $1$), for spin-$1/2$, yields 
a non trivial interplay between the dimensional crossover and the quantum fluctuation effects that induces a marked reduction of the interchain correlations. This effective reduction 
of the dimension due to magnetic frustration has been called {\it one dimensionalization}.\cite{Weihong06} Several analytical methods\cite{Balents10II} predict that such a quasi-one-dimensional regime persists until $J^{\prime}/J < 0.7$; however a variational Monte Carlo approach\cite{Heidarian09} predicts a sequence of continuous transitions: one at $J^{\prime}/J \sim 0.6$ from a  1D spin liquid phase to a two-dimensional (2D) spin liquid phase and another one at $J^{\prime}/J \sim 0.85$ to a 2D magnetic phase.

It is believed that this phenomenon is realized in the frustrated 2D magnetic compound Cs$_2$CuCl$_4$, 
where the broad continuum found in the spectrum measured by inelastic neutron scattering experiments was originally interpreted \cite{Coldea01} as a signal of 2D spinon excitation. 
Subsequent works \cite{Balents07, Balents10}, however, gave enough evidence of the 1D character of the  spinon excitations in agreement with the one dimensionalization scenario.
More recently \cite{Skoulatos17}, the dimensional reduction has been accomplished by controlling the pressure in the spin-$1/2$ magnetic material  CuF$_2$(D$_2$O)$_2$ (pyz) (pyz = pyrazine), allowing us to investigate the passage from spin wave to spinon excitation in the same triangular geometry.

On the other hand, for the spin-$1$ case the interplay between frustration and dimensional crossover has been little explored in the literature. On one side, numerical exact diagonalization studies \cite{Nakano13} do not allow estimating a reliable 
critical value due to the small system sizes investigated; on the other hand, using series expansion,\cite{Pardini08} Pardini and Singh estimated that the critical value 
between spiral magnetic and disordered phases is within the range $0.33<J^{\prime}/J<0.6$. However, the lack of an unbiased study of the short range correlations
did not allow discerning whether the effective reduction of the dimension actually occurs.  \\  

In this paper we investigate the phenomenon of one dimensionalization in the AF spin-$1$ Heisenberg model that interpolates between the 1D decoupled chains and the triangular lattice by mean of two complementary methods: density-matrix renormalization group (DMRG) and the Schwinger 
boson theory performed at the more reliable Gaussian correction level above the saddle-point solution. The DMRG results are very accurate for the regime of weakly coupled chains, while the Schwinger boson theory is suitable for the 2D magnetically ordered regime. The main DMRG result is that there is an abrupt transition between the decoupled spin chains and the spirally ordered 
regimes at $(J^{\prime}/J)_c\sim 0.42$ signaled by the sudden closing of the spin gap. This is in contrast to the unfrustrated case, where the gap closes continuously until long-range N\'eel order is set in. Coming from the 2D magnetically ordered side,  Schwinger boson theory predicts the instability of the spiral magnetic order at $(J^{\prime}/J)_c \sim 0.43$ toward a magnetically disordered phase with one-dimensional features. This agreement, along with 
the strong difference found with DMRG between the intra- and interchain short spin-spin correlations for sufficiently large values of the interchain coupling,    
allows us to confirm that the one-dimensionalization phenomenon is realized in the present frustrated spin-$1$ model. \\

This paper is organized as follows: in Section II we explain the model Hamiltonian. In Section III we describe the details of the DMRG method and the results. In Section IV we develop the Schwinger boson theory up to Gaussian correction level along with the results. In Section V we close with the conclusions.

\section{Model}

The spin-$1$ Heisenberg Hamiltonian that interpolates from 1D decoupled chains to the spatially isotropic square and triangular lattices can be written as 

\begin{equation}
H =  \sum_{ i } \left[ J\; \hat{\mathbf S}_i \hat{\mathbf S}_{i+\delta} + J' \;\hat{\mathbf S}_i \hat{\mathbf S}_{i+\delta'} + J^{\prime\prime}\; \hat{\mathbf S}_i \hat{\mathbf S}_{i+\delta''}\right],
\label{hamil}
\end{equation}

\noindent where the sum goes over all sites $i$ of the square lattice and the AF exchange interactions $J$, $J'$ and $J''$ connect the nearest-neighbor spins along the horizontal $\delta$ and diagonal $\delta'$ and $\delta''$ spatial directions, respectively, as shown in Fig. \ref{red}.  
Throughout this work we take $J=1$. As mentioned in the Introduction, the interpolation between decoupled spin chains and the square lattice is accomplished by varying  $J^{\prime}$ from $0$ to $1$  while keeping $J^{\prime\prime}=0$. On the other hand, to study the effect of frustration $J' = J''$ is assumed. So the interpolation between decoupled spin chains and the triangular lattice is accomplished by varying $J'$ from $0$ to $1$. 
 \\

\section{Density-Matrix renormalization group}

We use the standard density-matrix renormalization group algorithm \cite{White92} on square clusters $L \times L$ with cylindrical boundary conditions (periodic boundary conditions along the $J$ chains) for systems of $L= 4,6,8,10,12$ (see Fig. \ref{red}). We keep up to 1200 states for the worst case scenario ($J' = J''=1$ and large $L$) in order to keep the truncation error below $10^{-5}$, ensuring that errors become smaller than symbol size. In order to avoid the characteristic edge states of the Haldane chains (remember that for an $S=1$ open chain the spin gap should be calculated\cite{White92} as $\Delta = E_{S_z =2}- E_{GS}$) periodic boundary conditions are used along the $J$ chains. This allows us to calculate the spin gap as
\begin{equation}
\Delta = E_{S_z =1}- E_{GS}
\end{equation}
in interpolating between the decoupled chain and the 2D systems. The squared local magnetization is defined as 
\begin{equation}
 m^2 = \frac{S({\bf Q})}{N} = \frac{1}{N^2} \sum_{ij} \langle \hat{S}_i \hat{S}_j \rangle e^{i k (R_i-R_j)}
\end{equation}
\noindent where $S({\bf Q})$ is the structure factor evaluated at the magnetic wave vector ${\bf Q}$ and $N=L \times L$.\\
\begin{figure}[h]
\begin{center}
\includegraphics*[width=0.40\textwidth]{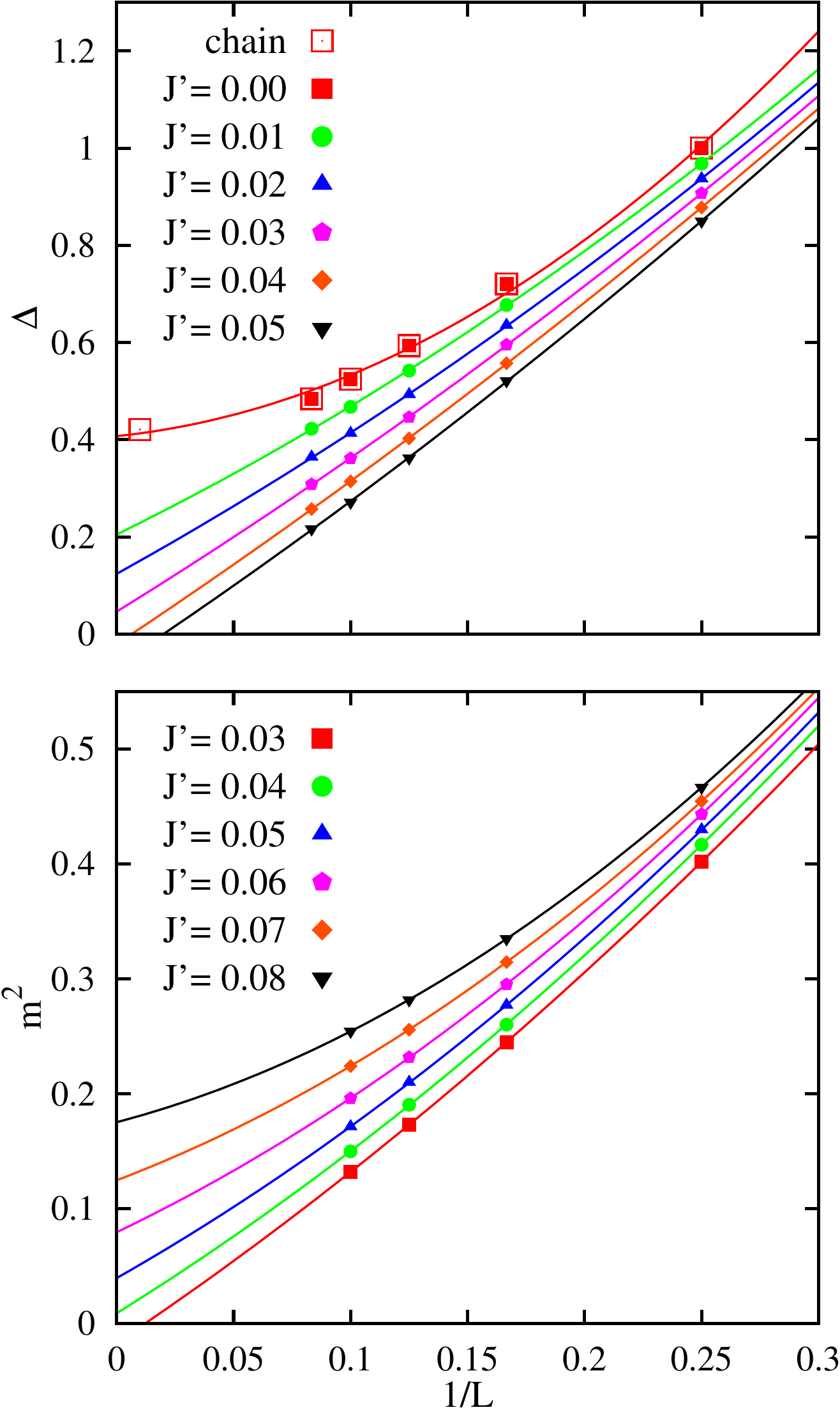}
\caption{DMRG results for the unfrustrated case $J^{\prime\prime}=0$: scaling of the spin gap $\Delta$ (top panel) and the squared staggered magnetization $m^2$ (bottom panel) versus $1/L$ for different values of the interchain coupling $J^{\prime}$. The system sizes shown are $L = 4,6,8,10$, and $12$. }
\label{gapm}
\end{center}
\end{figure}

\begin{figure}[t]
\begin{center}
\includegraphics*[width=0.40\textwidth]{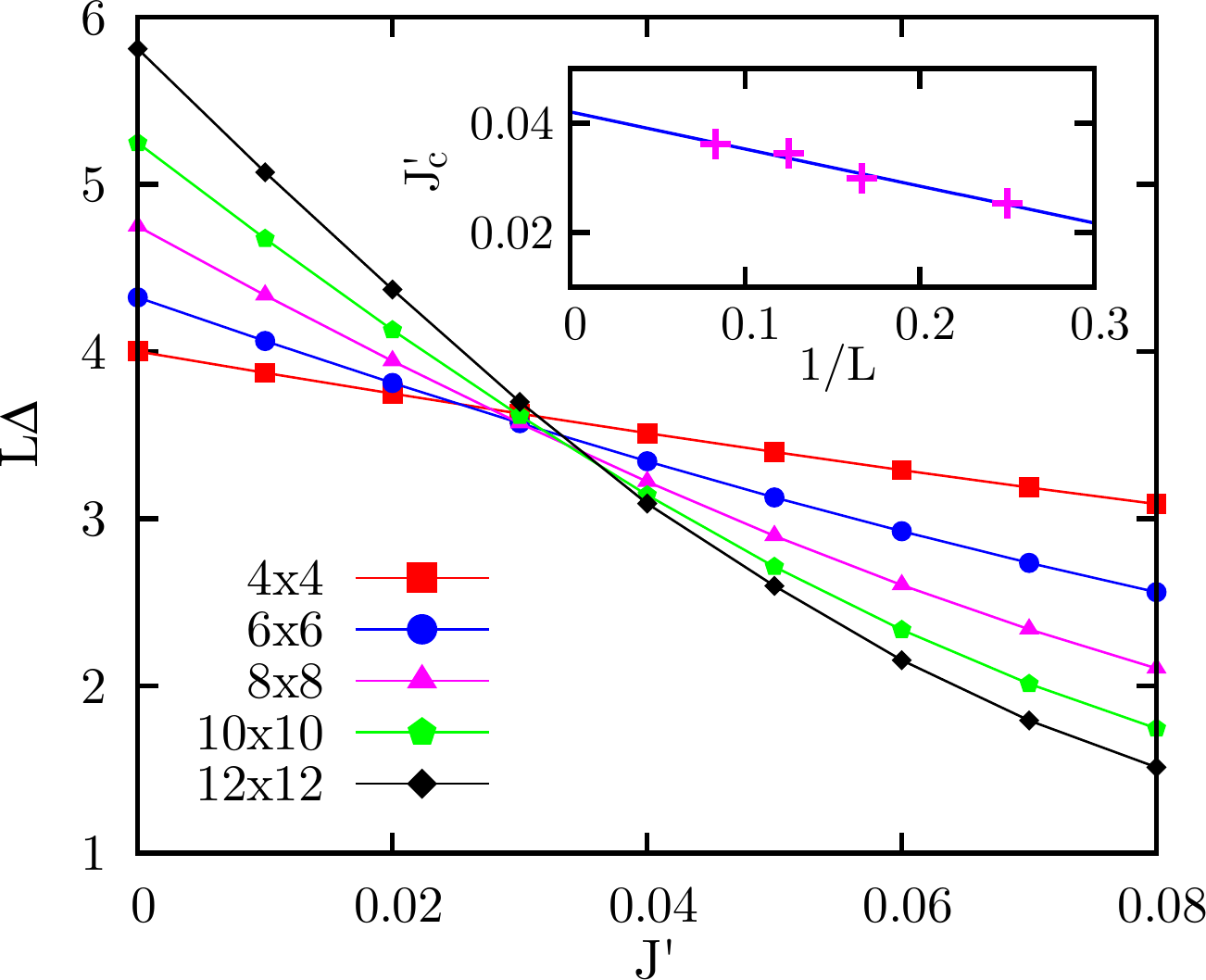}
\caption{DMRG results for the unfrustrated case $J^{\prime\prime}=0$: $L$ $\times$ $\Delta$ versus interchain coupling $J^{\prime}$ for different system sizes. Inset:
extrapolation of $J^{\prime}_c$ as the different crossing points for successive lattice sizes versus $1/L$. See the text.}
\label{gap_scal}
\end{center}
\end{figure}

In order to validate our DMRG method we have investigated the interpolation between the decoupled chain regime and the N\'eel order regime of the square lattice 
($J^{\prime\prime}=0$), studied previously.\cite{Matsumoto01,Moukouri11} In Figure \ref{gapm} shows the gap $\Delta$ (top panel) and the squared staggered magnetization
$m^2$ (bottom panel) versus $1/L$ for $L = 4,6,8,10,12$ and different values of the interchain coupling $J^{\prime}$. A quantum critical point is observed which 
is signaled by a continuous reduction of both the gap and the squared magnetization as the interchain coupling $J^{\prime}$ approaches a critical value $J^{\prime}_c$.
To quantitatively determine $J^{\prime}_c$ we use a finite-size scaling analysis over the gap $\Delta$ and the squared staggered magnetization $m^2$ as follows: near the quantum critical point the product $L\Delta$ (or $L m^2$) is given by a universal function, $L \Delta = f(C(J_c' - J') L^\frac{1}{\nu})$
where $L$ is the linear size of the system, $C$ is independent of $L$, and $\nu$ is the correlation length exponent.\cite{Moukouri11} When $J' = J^{\prime}_c$, $L\Delta = f(0)$ does not depend on the size $L$. Then, by assuming that the quantum critical point is scale invariant, a crossing of all curves $L\Delta$ versus $J^{\prime}$  at the critical interchain coupling $J^{\prime}_c$ (see Fig. \ref{gap_scal}) is expected. Due to the existence of finite-size effects such a crossing of the curves does not occurs. Therefore, in order to find $J^{\prime}_c$ in the thermodynamic limit, it is necessary to extrapolate the different crossing points for successive lattice sizes, as shown in the inset of Fig. \ref{gap_scal}. The value we find is $J^{\prime}_c= 0.042$, which agrees with the quantum Monte Carlo\cite{Matsumoto01} $J^{\prime}_c= 0.043648$ and 
two-step DMRG\cite{Moukouri11} $J^{\prime}_c= 0.043613$ predictions. Notice that by applying the above finite size procedure to the quantity $L m^2$ we find the same critical value. Even if the small value of $J^{\prime}_c$ with respect to the Haldane gap $\Delta\sim 0.41$ has been pointed out previously,\cite{Pardini08} why $J^{\prime}_c \ll \Delta$ has not been explained. Using heuristic arguments,   
the mean field energy gain due to $J^{\prime}$ coupling seems to be of order $zJ^{\prime} s^2 \xi $, where the  $\xi\sim6$ is the Haldane chain correlation length and 
$z$ is the number of neighboring chains. Then, at the quantum critical point  we find that $2J^{\prime}_c s^2\xi $ is of the same order of $\Delta$.\\

\begin{figure}[t]
\begin{center}
\includegraphics*[width=0.40\textwidth]{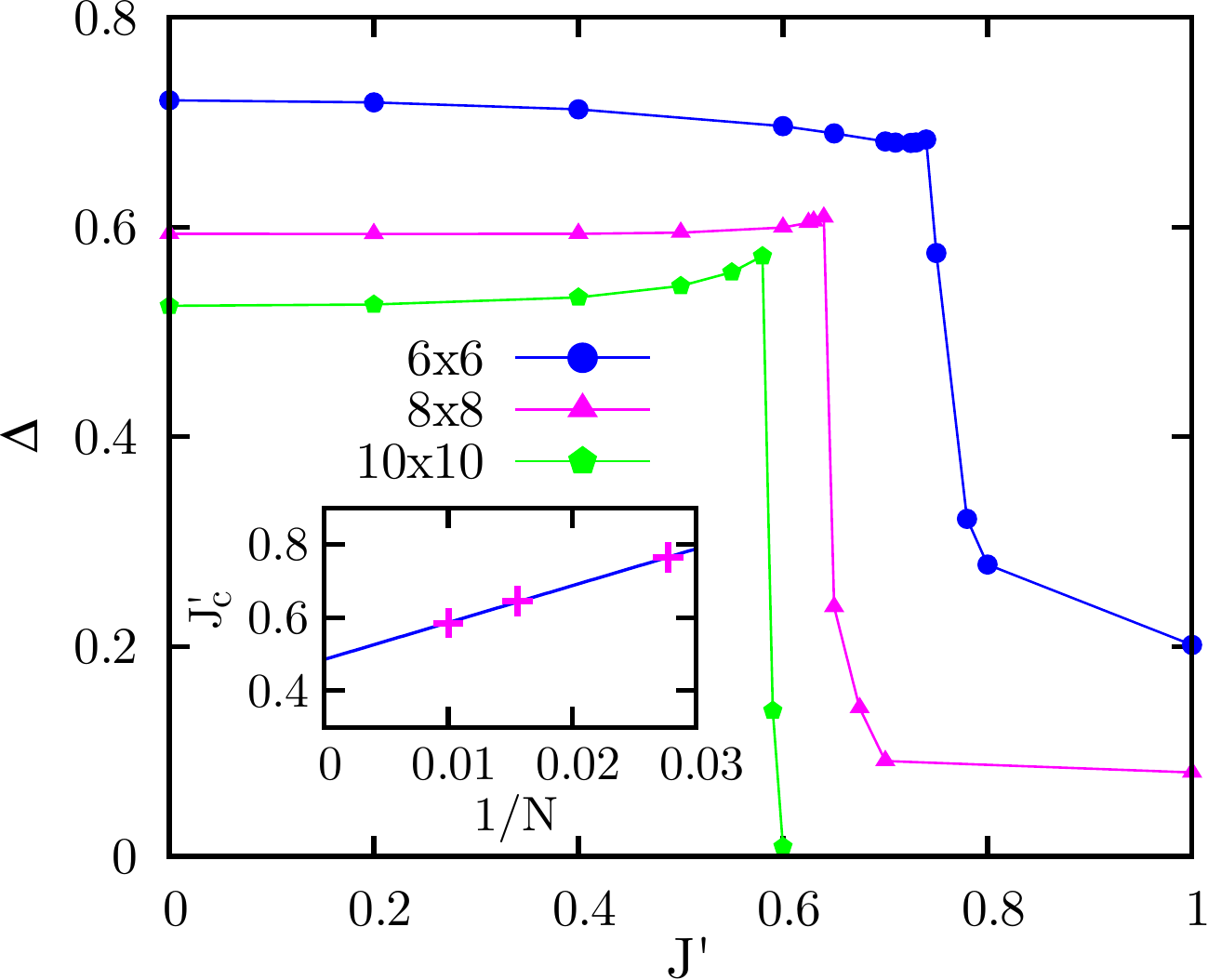}
\caption{DMRG results for the frustrated case $J^{\prime}=J^{\prime\prime}$: spin gap $\Delta$ versus interchain coupling $J^{\prime}$. The system sizes shown are $L = 6,8$, and $10$}
\label{gap_frustrated}
\end{center}
\end{figure}

\begin{figure}[h]
\begin{center}
\includegraphics*[width=0.40\textwidth]{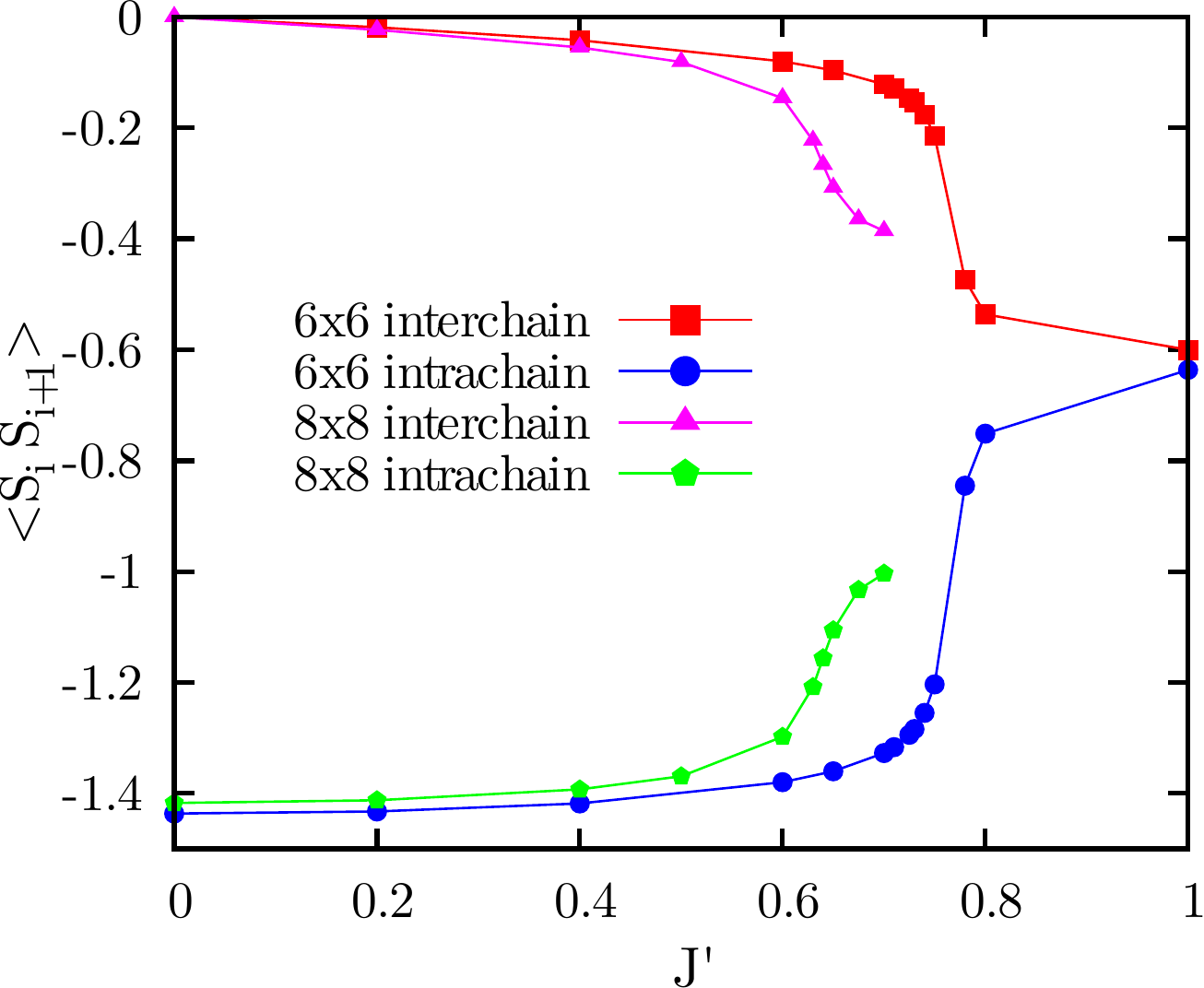}
\caption{DMRG results for the frustrated case $J^{\prime}=J^{\prime\prime}$: intra- and inter-chain spin-spin correlations between nearest neighbors as a function 
of $J^{\prime}$.}
\label{corr_frus}
\end{center}
\end{figure}

Once the DMRG procedure has been validated, we tackle the frustrated case, $J^{\prime}=J^{\prime\prime}$, where the behavior of the ground state with the interchain coupling is completely different from the unfrustrated case. 
In Fig. \ref{gap_frustrated} it is observed that the gap is practically unaltered for an important range of the interchain coupling. Furthermore, there is a sudden 
closing of the gap for each lattice size studied. The inset of Fig. \ref{gap_frustrated} plots the scaling of $J^{\prime}_c$ with $1/N$ where the critical value is assumed when the gap is halved for each size, being the extrapolated value $J^{\prime}_c=0.42$. Interestingly, this value  
is very close to the Haldane gap, in contrast to the unfrustrated case, where the interchain coupling is one order of magnitude smaller. Here, the previous mean-field arguments cannot be applied due to the frustration. 
Therefore, the magnetic frustration induced by the dimensional crossover seems to preserve the robustness of the Haldane phase. To investigate whether there is an effective reduction of the dimension we have studied the intra- and interchain short-range spin-spin correlations, which are shown in Fig. \ref{corr_frus}. Here, the quite abrupt change of the interchain correlations 
for large values of the interchain coupling resembles the behavior of the spin gap. For the $8\times8$ lattice we have not continued the calculation up to $J^{\prime}=1$ because the $120^{\circ}$ N\'eel order does not match this lattice size and, consequently, the isotropic regime is not recovered as in the $6\times6$ lattice. Although not shown in the figure, it should be noted that for the unfrustrated case 
the interchain short range spin-spin correlations increase smoothly as a function of  $J^{\prime}$.

Unfortunately, coming from the 2D magnetically ordered regime, the DMRG computation of the critical value $J^{\prime}_c$  is quite difficult 
to determine since incommensurate spiral phases are expected near the transition to the Haldane regime. This requires the implementation 
of open boundary conditions along with
larger lattice sizes in order to properly accommodate  the magnetic wave vector\cite{Weichselbaum11}. Given this difficulty, in the next section, we resort to 
the Schwinger boson theory, which is reliable for the study of 2D systems.\cite{Ceccatto93,Trumper97,Manuel99,Flint08}

\section{Schwinger boson theory }
In the Schwinger boson theory
the spin operator is represented as 
${\bf S}_i= \frac{1}{2}{\bf b}^{\dagger}_i \vec{\sigma} \;{\bf b}_i$, with $\vec{\sigma}$ being the Pauli matrices and the spinor
${\bf b}^{\dagger}_i =(\hat{b}^{\dagger}_{i\uparrow}; \hat{b}^{\dagger}_{i\downarrow})$
composed of the Schwinger spin-$\frac{1}{2}$ bosons $\hat{b}_{\uparrow}$ and $\hat{b}_{\downarrow}$, subject to the local 
constraint $\sum_{\sigma}\hat{b}^{\dagger}_{i\sigma}\hat{b}_{i\sigma}=2s$. Then, Eq. (\ref{hamil}) can be rewritten as
\begin{equation}
H = \sum_{<i,j>} J_{ij} \left( :\hat{B}_{ij}^\dagger \hat{B}_{ij} : - \hat{A}_{ij}^\dagger \hat{A}_{ij} \right),
\label{AABB}
\end{equation}
where the link operators $\hat{B}_{ij}^\dagger\!\!\!\!=\!\!\!\frac{1}{2} \sum_\sigma \hat{b}_{i\sigma}^\dagger \hat{b}_{j\sigma}$ and 
$\hat{A}_{ij}\!\! =\!\!\frac{1}{2} \sum_\sigma \sigma \hat{b}_{i\sigma} \hat{b}_{j\bar{\sigma}}$ are  $SU(2)$ invariant and
$J_{ij}$ takes the values $J, J^{\prime},J^{\prime\prime},$ depending on the link direction $ij=\delta,\delta',\delta''$, respectively, as shown in Fig. \ref{red}.
\\

The partition function is written as the functional integral over coherent states of Schwinger bosons\cite{Auerbach94}
\begin{multline}
Z =  \int D\lambda [D\overline{b}D{b}] \ e^{-\int_0^\beta d\tau \left[ \sum_{i,\sigma} \overline{b}_{i,\sigma}^\tau \partial_\tau b_{i,\sigma}^\tau + H(\overline{b},b) \right] } \\
 \times e^{-\int_0^\beta d\tau \left[ i \sum_i \lambda_i^\tau (\sum_\sigma \overline{b}_{i,\sigma}^\tau b_{i,\sigma}^\tau -2S )\right] } ,
\end{multline}
where the boson operators are replaced by complex variables, the $\lambda$ field is added to enforce the local constraint over the number of bosons per site,	 
and the 
integral measures are defined as $[D\overline{b}D{b}]=\Pi \frac{d\overline{b}^{\tau}_{i,\sigma}d{b}^{\tau}_{i,\sigma}}{2\pi i}$ and  $D\lambda =\Pi \frac{d\lambda^{\tau}_i}{2\pi i}$.  Two types of Hubbard-Stratonovich transformations are introduced \cite{Trumper97,Flint08} to decouple the $\hat{B}_{ij}^\dagger \hat{B}_{ij}$ and $\hat{A}_{ij}^\dagger \hat{A}_{ij}$ terms of Hamiltonian (\ref{AABB}), so the partition function can be re-written as
\begin{equation}
Z = \int D\overline{W}DW D\lambda \ e^{-S_{\text{eff}}(\overline{W},W,\lambda)},
\label{exact}
\end{equation} 

\noindent where $D\overline{W}DW$ denotes the measure of the new Hubbard-Stratonovich complex fields 
$\overline{W}^{A \tau}_{ij}$, $\overline{W}^{B \tau}_{ij}$,
$W^{A \tau}_{ij}$, and $W^{B \tau}_{ij}$ which depend on site $i$, imaginary time $\tau$, link direction $ij$,   
and field index $A$, $B$ according to the term they are decoupling in Eq. (\ref{AABB}). The effective action $S_{\text{eff}}$ is given by 
    
\begin{multline}
S_{\text{eff}}\! =\!\! \int^{\beta}_0 \!\!\! d\tau (\sum_{i,j,\mu}\!\! J_\delta \overline{W}_{ij}^{\mu,\tau} W_{ij}^{\mu,\tau} - i2S \sum_i \lambda_i^{\tau}) - \ln Z_{\text{bos}},
\end{multline}
where $\mu$ sums over the $A$, $B$ index fields and $Z_{\text{bos}}$ is the bosonic partition function 

\begin{equation}
 Z_{\text{bos}}= \int [D\overline{b}D{b}] e^{-S_{\text{bos}}(\overline{b},{b})}
\end{equation}

\noindent with the resulting quadratic bosonic action given by
\begin{equation}
S_{\text{bos}} = \int^{\beta}_{0} \!\!\! d\tau \sum_{i,j} \vec{b}_i^{\tau\dagger} \mathcal{M}^{\tau}_{i,j} \vec{b}^{\tau}_{j},
\end{equation}
\noindent where  $\vec{b}_i^{\tau\dagger} = (\overline{b}^{\tau}_{i \uparrow}, {b}_{i \downarrow}^{\tau})$ and $\mathcal{M}^{\tau}_{i,j}$ is the dynamical matrix defined as \\

 {\small $\mathcal{M}^{11}_{ij}= (\partial_{\tau}+i \lambda _i) \delta_{ij}+\frac{J_{ij}}{2} (W^{B}_{ij}-\overline{W}^{B}_{ji});\;\;\;\;\;{\mathcal{M}^{22}_{ij}}=-\overline{\mathcal{M}^{11}_{ij}}$}, \\

{\small $\mathcal{M}^{12}_{ij}= \frac{J_{ij}}{2} (W^{A}_{ji}-{W}^{A}_{ij});\;\;\;\;\;\;\;\;\;\;\;\;\;\;\;\;\;\;\;\;\;\;\;\;\;\;\;\;\;\; \mathcal{M}^{21}_{ij}=\overline{\mathcal{M}^{12}_{ij}}$}. \\

\noindent So far the formulation of the partition function $Z$ is exact. To compute approximately Eq. (\ref {exact}) the effective action $S_{\text{eff}}$ is expanded around the saddle-point solution of the fields. Going to frequency and momentum space, the effective action to quadratic order results in

\begin{equation}
S_\text{eff} \approx  S^{0}_\text{eff} + \frac{1}{2} \sum_{\alpha_1,\alpha_2} \Delta \vec{\phi}_{\alpha_1}^\dagger S^{(2)}_{\alpha_1,\alpha_2}  \Delta \vec{\phi}_{\alpha_2}, 
\end{equation}

\noindent where $S^{0}_\text{eff}$ and $S^{(2)}_{\alpha_1,\alpha_2}=\frac{\partial^2 S_\text{eff}}{\partial \vec{\phi}_{\alpha_1}^\dagger \partial \vec{\phi}_{\alpha_2}}$
are the effective action and the fluctuation matrix, respectively, both evaluated  at the saddle point solutions; while $\Delta \vec{\phi}_\alpha^\dagger = \vec{\phi}_\alpha^\dagger - \vec{\phi}_\text{sp}^\dagger $ 
are the fluctuations of the fields $\vec{\phi}_\alpha = (W_{\alpha \delta}^B,\overline{W}_{-\alpha \delta}^B,W_{\alpha \delta}^A,\overline{W}_{-\alpha \delta}^A,
\lambda_\alpha)^T$ around the saddle point solutions $\vec{\phi}_\text{sp}^\dagger$, 
which are assumed to be static and homogeneous. Notice that each link field has components along the $\delta=\delta,$ $\delta'$, and $\delta''$ directions and  $\alpha= {\bf k},\omega$.
The partition function (\ref {exact}) thus approximated consists of carrying out the Gaussian integral over the fluctuation fields (see below). In principle, this approximation is valid for a large number of   
Schwinger bosons flavors $N$.\cite{Auerbach94} In fact, for $N=\infty$ the saddle point solution is exact. 
However, for $N=2$ it has already been shown that such Gaussian corrections notably improve the saddle-point solution in related Heisenberg models \cite{Trumper97,Manuel99}. 

\subsection{Saddle-point approximation}
The saddle point solution is found by solving

\begin{equation}
{\frac{\partial S_{\text{eff}}}{\partial \vec{\phi}_{\alpha}}}  =  \vec{\psi}_\alpha^\dagger  - 
Tr \left( \mathcal{G}_{sp} v_{\alpha} \right) = 0,
\label{SP}
\end{equation}

\noindent where {\small $\vec{\psi}_\alpha^\dagger \!\!=\!\! {(J \overline{W}_{\alpha \delta}^B,J {W}_{-\alpha \delta}^B,J \overline{W}_{\alpha \delta}^A,J {W}_{-\alpha \delta}^A,-i2S(N\beta)^{\frac{1}{2}}\delta_{\alpha,0})}$}; the trace goes over momentum ${\bf k}$, frequency $\omega$, and the bosonic flavor index; 
$v_{\alpha}\!\!=\!\!{\partial \mathcal{M}}/{\partial \vec{\psi}_{\alpha}}$; and $\mathcal{G}_\text{sp}$ is the saddle point Green's function, defined as $\mathcal{G}_\text{sp} = \mathcal{M}_\text{sp}^{-1}$ with 

\begin{equation}
\mathcal{M}_{sp}  = \left( \begin{array}{cc}
[i\omega+\lambda+\gamma_{\bf k}^B] & -\gamma_{\bf k}^A \\
-\gamma_{\bf k}^A & [-i\omega+\lambda+\gamma_{\bf k}^B]
\end{array} \right) ,
\end{equation}

 \noindent {\small $\gamma_{\bf k}^B \!=\! \sum_{\delta} J_\delta B_\delta \cos({\bf k}\!\cdot\!\delta)$}, and  {\small
 $\gamma_{\bf k}^A \! =\! \sum_{\delta} J_\delta {A}_\delta \sin({\bf k}\!\cdot\!\delta)$}. Here ${A}_\delta$, ${B}_\delta$ and $\lambda$ are the mean field parameters which are chosen to be real and related to the saddle point fields as 

\begin{equation}
\begin{array}{ccc}
\left. \overline{W}_{{\bf 0},\delta}^B \right|_\text{sp} = ({N\beta})^{\frac{1}{2}}{B}_\delta &   & \left. \overline{W}_{{\bf 0},\delta}^A \right|_\text{sp} = -i({N\beta})^{\frac{1}{2}}{A}_\delta \\

\left. {W}_{{\bf 0},\delta}^B \right|_\text{sp} = -({N\beta})^{\frac{1}{2}}{B}_\delta &   & \left. {W}_{{\bf 0},\delta}^A \right|_\text{sp} = i({N\beta})^{\frac{1}{2}}{A}_\delta\\

& \left. \lambda \right|_\text{sp}=i \lambda&

\end{array}
\end{equation}

\noindent As a consequence of the sign difference \cite{Trumper97,Flint08} of each term in Eq. (\ref{AABB}) $\left. \overline{W}_{{\bf 0},\delta}^B \right|_\text{sp} \!\!\!\!=\!\!-\left. {W}_{{\bf 0},\delta}^B \right|_\text{sp}$; whereas $\left. \overline{W}_{{\bf 0},\delta}^A \right|_\text{sp} =
\left. ({W}_{{\bf 0},\delta}^A \right|_\text{sp})^{*}.$ The poles of $\mathcal{G}_\text{sp}$ correspond to the free spin-$1/2$ spinon excitation

\begin{equation}
\varepsilon_{\bf k} = \sqrt{\left(\gamma_{\bf k}^B+\lambda \right)^2-\left(\gamma_{\bf k}^A\right)^2},
\end{equation}

\noindent which is usually found within the Schwinger boson mean-field theory \cite{Ceccatto93,Mezio11}. The self-consistent equations, resulting from equation (\ref{SP}), 
have the well-known  zero-temperature form 
\begin{equation}
{A}_\delta = \frac{1}{2N} \sum_{\bf k} \frac{\gamma_{\bf k}^A}{\varepsilon_{\bf k}} \sin({\bf k}.\delta) 
\end{equation}
\begin{equation}
{B}_\delta = \frac{1}{2N} \sum_{\bf k} \frac{\gamma_{\bf k}^B + \lambda}{\varepsilon_{\bf k}} \cos({\bf k}.\delta)
\end{equation}
\begin{equation}
S+\frac{1}{2} = \frac{1}{2N} \sum_{\bf k} \frac{\gamma_{\bf k}^B + \lambda}{\varepsilon_{\bf k}}.
\label{self}
\end{equation}

\noindent The solutions of the above saddle point equations correspond to a singlet ground state \cite{Mezio11}. However, as the system size $N$ increases the ground state develops magnetic correlations signaled by the minimum gap of the spinon dispersion located at momenta $\pm{{\bf Q}_0}/{2}$, where ${\bf Q}_0$ is the magnetic wave vector that varies according to the values of $J, J^{\prime}$, and $J^{\prime\prime}$. In two dimensions the spinon gap may behave as $\varepsilon_{\pm{{\bf Q}_0}/{2}}\sim 1/N$. 
In this case, for large system sizes, the zero modes can be treated as a Bose condensation which is interpreted as the rupture of the $SU(2)$ symmetry \cite{Hirsch89,Chandra90}. Usually, the presence of long range order is described by the local magnetization $m({\bf Q}_0)$. Alternatively, one can compute the spin stiffness in the following way \cite{Manuel99}: the self-consistent 
equations are solved with twisted boundary conditions so as to get the saddle-point solution corresponding to a magnetic structure slightly twisted an amount 
$\Delta {\bf Q}$ 
with respect to the periodic boundary conditions case ${\bf Q}_0$. For different values of the exchange coupling the saddle-point ground-state energy takes the form    
{\small
\begin{equation}
E_{\scriptscriptstyle SP}({\bf Q})\! =\! N\!\left[ J (B^2_{\delta}\!-A^2_{\delta})\!+\!J' (B^2_{\delta'}\!-A^2_{\delta'})\!+
                                       \!J'' (B^2_{\delta''}\!-\!A^2_{\delta''}) \right], 
\label{ESP}
\end{equation}
}

\noindent where ${\bf Q}$ also depends on the twisted boundary conditions imposed. Then, the saddle point spin stiffness $\rho_{\scriptscriptstyle SP}$ is obtained  numerically  by computing $\rho_{\scriptscriptstyle SP}=\partial^2 E_{\scriptscriptstyle SP}({\bf Q})/\partial {\bf Q}^2$. The reason why we focus 
on the spin stiffness, instead of the local magnetization, is because the ground-state energy $E({\bf Q})$ is easier to compute to Gaussian order, giving
us access to the Gaussian corrections of $\rho_{\scriptscriptstyle SP}$ (see below).

\subsection{Gaussian fluctuations}
The fluctuation matrix evaluated at the saddle-point solution can be written as
\begin{equation}
S^{(2)}_{\alpha_1,\alpha_2} = \frac{\partial S_{eff} }{\partial\vec{\phi}^{\dagger}_{\alpha_1} \partial\vec{\phi}_{\alpha_2} }= \frac{\partial \vec{\psi}_{\alpha_2}^\dagger}{\partial \vec{\phi}_{\alpha_1}^\dagger} - Tr \left[  \mathcal{G}_{sp} v_{\alpha_2} 
 \mathcal{G}_{sp} v_{\alpha_1}
\right].
\label{S2}
\end{equation}
 
\noindent The computation to Gaussian order of the partition function (\ref{exact}) implies carrying on the Gaussian integral 

\begin{equation}
Z\cong e^{-S^{0}_\text{eff}}  \int D\vec{\phi}^{\dagger} D\vec{\phi}  \ e^{- \frac{1}{2} \Delta \vec{\phi}^\dagger S^{(2)}  \Delta \vec{\phi}}.
\label{gauss}
\end{equation}

\noindent However, some care must be taken into account since, due to the rupture of the local gauge symmetry of the saddle-point solution, $S^{(2)}$ has infinite zero-mode gauge fluctuations that lead to divergences. To avoid them we introduce the Fadeev-Popov trick which 
restricts the integration to field fluctuations orthogonal to the gauge orbit \cite{Trumper97}. This procedure gives  the 
following Gaussian correction of the free energy:

\begin{equation}
 {F}^{(2)}=\frac{-1}{2\beta} \sum_{{\bf k} \omega_n} \ln \left[ \frac{\Delta_{FP}({\bf k},i\omega_n)}{\omega^2_n det S_{\perp}({\bf k},i \omega_n)} \right],
\end{equation}

\noindent where $\Delta_{FP}({\bf k},i\omega_n)= 8 \sum_{\delta} [(1+\cos {\bf k} \cdot \delta) A^2_{\delta} - (1-\cos {\bf k} \cdot \delta) B^2_{\delta}- (i\omega_n)^2]$ 
is the Fadeev-Popov determinant and $S_{\perp}({\bf k},i\omega_n)$ is the projection of the fluctuation matrix onto the subspace orthogonal by the right to the zero gauge modes. 
At $T=0$, the Gaussian correction to the ground state energy is

\begin{equation}
 E^{(2)}=\frac{-1}{4\pi N} \int^{\infty}_{-\infty}d\omega \sum_{\bf k} \ln \left[ \frac{\Delta_{FP}({\bf k},\omega)}{\omega^2 det S_{\perp}({\bf k},\omega)} \right].
\label{E2}
\end{equation}

\noindent Therefore, the ground state energy for any twisted boundary conditions turns out to be 

\begin{equation}
 E_{\scriptscriptstyle FL}({\bf Q})=E_{\scriptscriptstyle SP} + E^{(2)},
\label{EGAUSS}
\end{equation}

\noindent where the Gaussian correction to the saddle point spin stiffness can be computed as  $\rho_{\scriptscriptstyle FL}=\partial^2 E_{\scriptscriptstyle FL}({\bf Q})/\partial {\bf Q}^2$.

\subsection{Schwinger boson results}
\begin{figure}[t]
\begin{center}
\includegraphics*[width=0.4\textwidth]{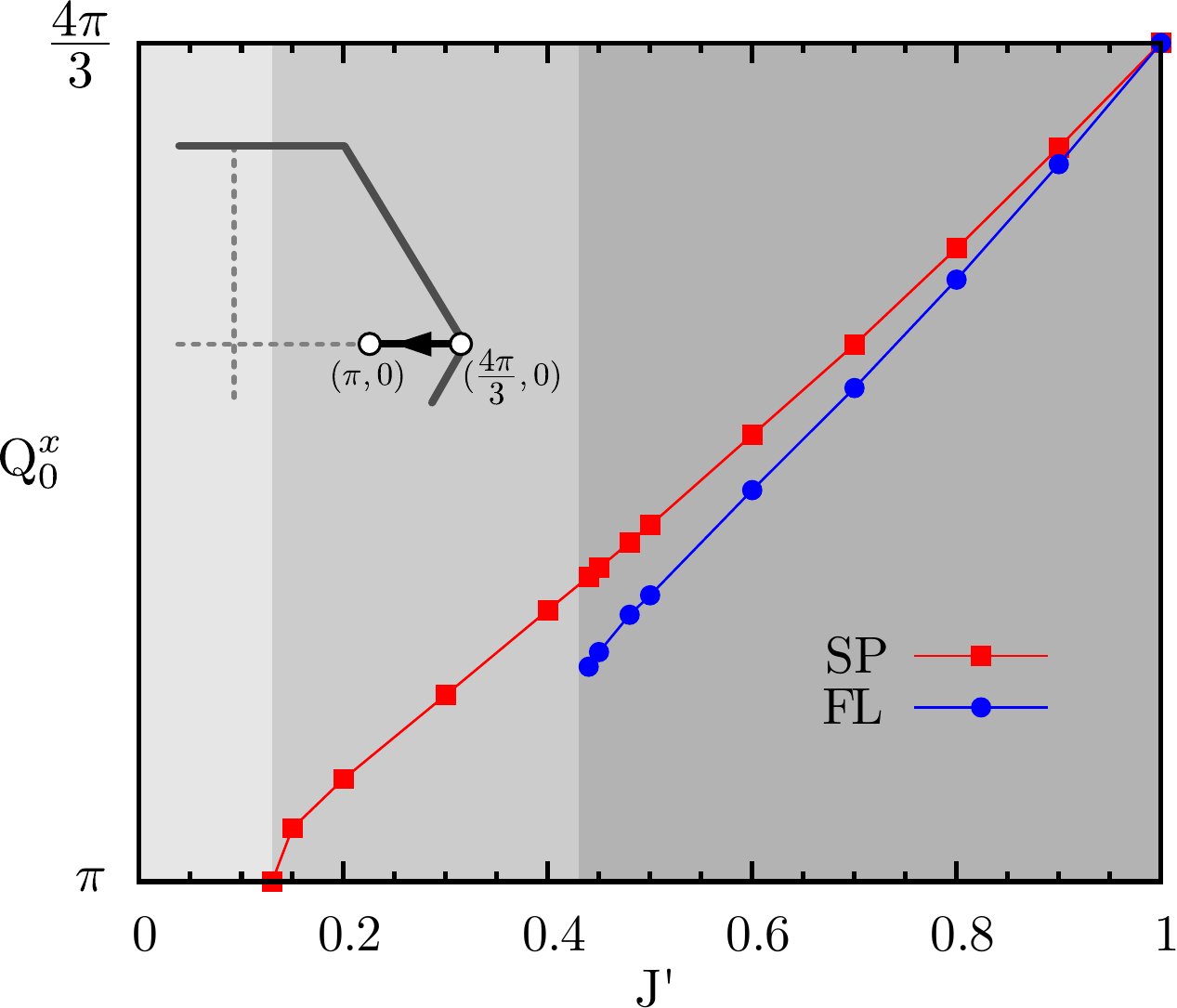}
\caption{Magnetic wave vector $Q_0$ predicted by the Schwinger bosons as a function of $J^{\prime}=J^{\prime\prime}$. Red squares and blue circles
 correspond to the saddle-point solution  and Gaussian corrections, respectively. The evolution of ${\bf Q}_0$ from the isotropic triangular to the 
decoupled chain regime is shown in the top left (see the text).}
\label{wavevector}
\end{center}
\end{figure}

In this section we analyze the Schwinger boson results going from the triangular isotropic case $J=J^{\prime}=J^{\prime\prime}$ to  the
completely decoupled chain case $J^{\prime}=J^{\prime\prime}=0$. Semiclassically, this implies a change of the magnetic wave vector ${\bf Q}_0$ 
from $(\frac{4\pi}{3},0)$ to $(\pi,0)$, via intermediate incommensurate spiral values (see top left in Fig. \ref{wavevector}). Actually, in the regime of decoupled chains there are many degenerate 
ground states which can be designated as $(\pi,q)$, meaning that the spin-spin correlations are N\'eel-like along the chains whereas the interchain-spin spin correlations are arbitrary. Figure \ref{wavevector} shows the magnetic wave vectors 
for different values of the exchange couplings predicted 
by the Schwinger boson theory. The saddle-point solutions (red squares) and the Gaussian corrections (blue circles) are obtained from the minima of Eqs. (\ref{ESP}) and (\ref{EGAUSS}), respectively. Except for the isotropic case where the magnetic wave vector ${\bf Q}_0=(\frac{4\pi}{3},0)$ is maintained after Gaussian corrections,
for the anisotropic cases the magnetic wave vectors of the incommensurate spiral orders are renormalized.  
 At the saddle-point level the transition to the decoupled chain regime, ${\bf Q}_0=(\pi,0)$, occurs at the critical value $J_c' = 0.13$, while Gaussian corrections 
render the incommensurate spiral orders unstable for $0.13 < J^{\prime} < 0.43$. In particular, these instabilities occur for a given ${\bf k}$ and $\omega=0$ where the determinant of    
$S_{\perp}({\bf k},\omega)$ [in Eq. (\ref{E2})] becomes negative. In these cases $E^{(2)}$ can not be calculated anymore. Remarkably, the value $J^{\prime}_c=0.43$ is very close to $J^{\prime}_c = 0.42$ found with DMRG 
(Sec. III). In addition, within the range  $0<J^{\prime}< 0.43$ the weakly coupled chains regime is stable after Gaussian corrections. This precludes the possibility of an intermediate 2D spin liquid state as in the spin-$\frac{1}{2}$ case, although it must be stressed that the weakly 
coupled chain regime is not well described by the Schwinger bosons theory. \\

\begin{figure}[t]
\begin{center}
\includegraphics*[width=0.4\textwidth]{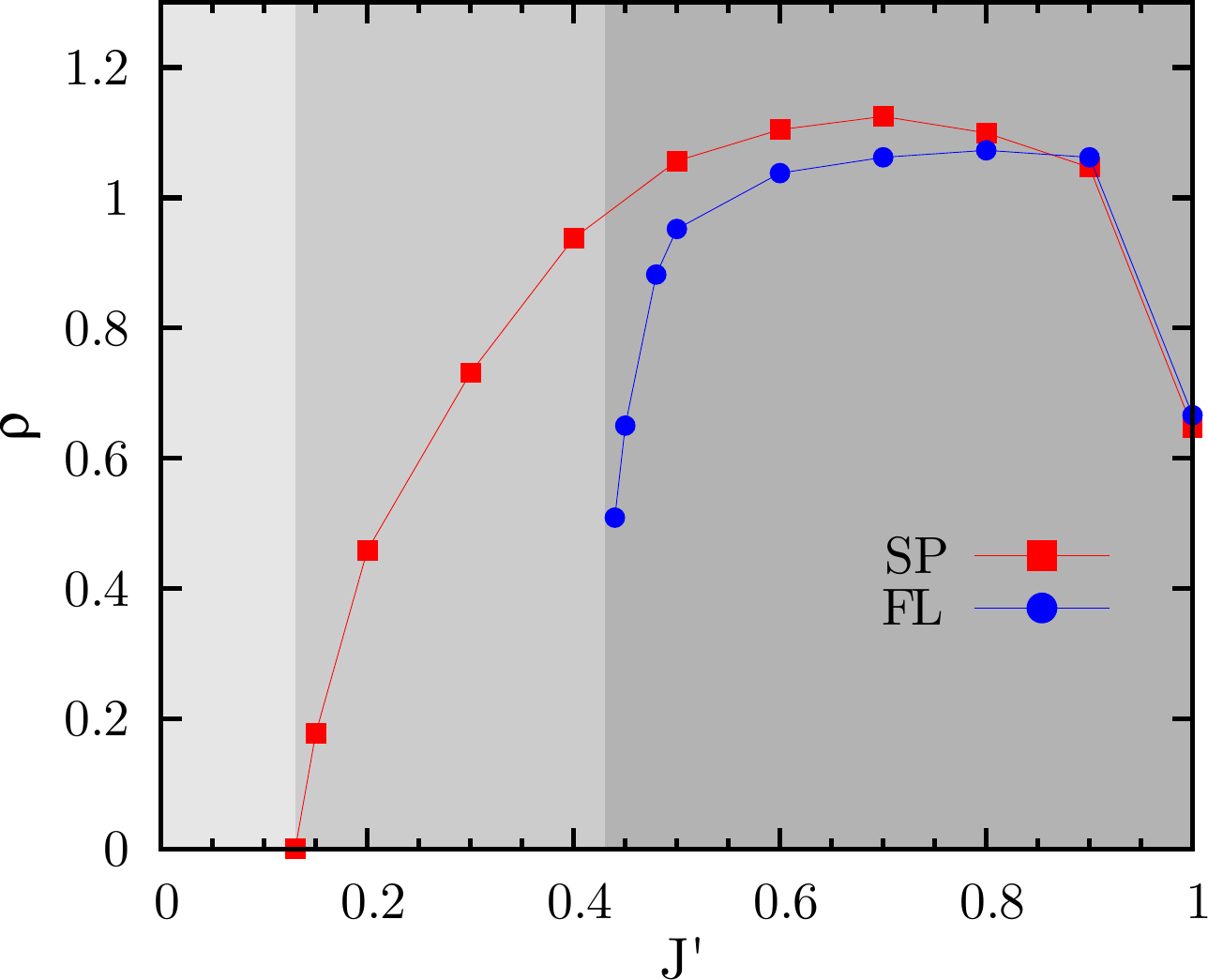}
\caption{Spin stiffness predicted by the Schwinger bosons as a function of $J^{\prime}=J^{\prime\prime}$. 
Red squares and blue circles correspond to the saddle-point solution and Gaussian corrections, respectively.}
\label{stiff}
\end{center}
\end{figure}

\noindent Figure \ref{stiff} shows the spin stiffness within the saddle-point approximation (red squares) along with the Gaussian corrections (blue circles). As explained previously, the saddle-point spin stiffness is obtained by first solving the self-consistent equations with twisted boundary conditions, then 
plugging in the twisted mean-field parameters in Eq. (\ref{ESP}), and finally computing numerically the second order derivative of the ground-state energy $E_{\scriptscriptstyle SP}$ with respect to ${\bf Q}$. On the other hand, the Gaussian corrected spin stiffness is obtained by deriving $E_{FL}$ 
once the twisted mean-field parameters are plugged in Eq. (\ref{E2}). Coming from the isotropic case $J^{\prime}=1$ the Gaussian corrections for the spiral phases tend to weaken the 
spin stiffness until the ground state becomes unstable at     
$J^{\prime}_c=0.43$, in accordance with Fig. \ref{wavevector}. Furthermore, there is an abrupt transition to a magnetically disordered phase with 1D character which is in line with the DMRG results in Sec. III. 
Based on these results, in particular on the behavior of the DMRG gap results, we can conclude that there is no room for an intermediate 2D spin liquid phase and that, in contrast to the spin-$1/2$ case, the one dimensionalization phenomenon for the spin-$1$ case occurs quite abruptly near $ J_c^{\prime}=0.42$. \\

\noindent To complete our Schwinger boson study we report the results for the anisotropic square lattice case ($J^{\prime\prime}=0$).    
At the saddle-point, coming from the isotropic square-lattice, the spin stiffness vanishes at $J_c'=0.0092$, which matches previous studies \cite{Ji03}. 
This value is far from the numerical\cite{Matsumoto01,Moukouri11,Wierschem14} results $J_c'=0.043648$. As in the frustrated case, this difference can be attributed to the tendency of the Schwinger boson mean field to favor magnetically ordered phases. The Gaussian corrected spin stiffness, however, vanishes at $J_c' = 0.0265$, getting closer to the above numerical results. \\

\section{Conclusions}
We have investigated the one dimensionalization phenomenon in the spin-$1$ Heisenberg model on the anisotropic triangular lattice using two complementary
methods: DMRG and Schwinger boson theory computed up to Gaussian correction level. Based on the ability of the DMRG method  and the Schwinger boson theory to give reliable 
results in interpolating from the decoupled chain and the 2D regimes, respectively, we can conclude that the effective reduction of the dimension occurs abruptly
near $J^{\prime}_c=0.42$. This value is very close to the Haldane gap  and is in contrast to the unfrustrated case where the  critical value is one order of magnitude smaller than the spin gap. Finally, even if the one-dimensionalization  phenomenon was first observed in the spin-$1/2$ case, 
we can conclude that it is not necessarily related to the critical nature of the spin chain ground state but to the interplay between the frustration and the dimensional crossover, which in the spin-$1$ case seems to, effectively, promote the  reduction of the dimension.\\

\noindent More recently, the one-dimensionalization phenomenon in spin-$1$ systems, has been studied \cite{Niesen17} in the bilinear-biquadratic Heisenberg model on the square lattice where an intermediate Haldane phase has been found in a narrow range of parameter space 
between the usual N\'eel state and three sublattice states forming $120^{\circ}$. We hope the present work can help future studies of the one-dimensionalization phenomenon in spin-$1$ systems.\\     

\section*{Acknowledgment}

\noindent This work was supported by CONICET under Grants  No. 364 (PIP2015) and No. 389 (PIP2012).

\end{document}